\documentclass[12pt, preprint]{aastex}
\newcommand{\siml}{\lower4pt \hbox{$\buildrel < \over \sim$}}
\newcommand{\simg}{\lower4pt \hbox{$\buildrel > \over \sim$}}
\def\paren#1{\left( #1 \right)}

\def\bra#1{\left[ #1 \right]}

\begin{document}
\title{Gravitational Wave and X-ray Signals from Stellar 
Disruption by a Massive Black Hole}

\author{Shiho Kobayashi\altaffilmark{1,2}, 
Pablo Laguna \altaffilmark{1,2},
E. Sterl Phinney\altaffilmark{3},
Peter M\'esz\'aros \altaffilmark{1,2,4} }

\altaffiltext{1}{Dpt. of Physics and Center for Gravitational Wave Physics,
   Pennsylvania State University, University Park, PA 16802}
\altaffiltext{2}{Dpt. of Astronomy \& Astrophysics, Pennsylvania State
   University, University Park, PA 16802}
\altaffiltext{3}{California Institute of Technology, Pasadena, CA 91125}
\altaffiltext{4}{Institute for Advanced Study, Princeton, NJ 08540}
\begin{abstract}
Gravitational waves and X-ray flares are expected from tidal
disruption of  stars by a massive black hole. Using a relativistic 
smoothed particle hydrodynamics code, we investigate the fate of 
main sequence and Helium stars in plunge orbits passing near 
Schwarzschild or a Kerr black holes of  mass $\sim 10^{5-6} M_\odot$. 
We show that quadrupole gravitational waves emitted during the
tidal disruption process are described reasonable well by a point
particle approximation even in the strong encounter case. 
An additional hydrodynamic calculation based on the Godunov method 
indicates that shocks develop for sufficiently high tidal compressions.
The shock-heating results in an X-ray flare, which for solar-type 
stars disrupted by $\sim 10^6M_\odot$ black holes is in the keV range, 
associated with the gravitational wave signal. The hardness and duration
of the X-ray flare may serve as a diagnostic of the 
mass of the central black hole. 

\end{abstract}
%
\section{Introduction}

There is strong evidence, based on stellar kinematics and gas 
dynamics, for the existence of massive black holes with masses from 
a few times $10^5$ to a few times $10^9 M_\odot$ in the center of a
substantial fraction of galaxies (e.g. Kormedy \& Gebhardt 2001 ). 
Gravitational radiation caused by stars orbiting closely around
such black holes is one of the most promising candidates for 
detection by the Laser Interferometer Space Antenna (LISA; e.g.
Danzmann et al. 1998; Hils \& Bender 1995; Sigurdsson \& Rees 1997; 
Ivanov 2002; Freitag 2003). Compact stars (white dwarfs, neutron stars,
or stellar mass black holes) in bound orbits relatively close to the 
black hole will emit gravitational waves as they spiral all the 
way down to the horizon. Although the detection of such waves requires
a set of accurate wave templates and significant computational power 
to search the large dimensionality parameter space, tracking the 
signals over a large number of cycles boosts the strength of the signals, 
hence previous studies have mainly addressed the gravitational waves 
from such stars.

On the other hand, there is a larger reservoir of main sequence 
stars, which are affected by the gravitational field of the  massive black 
hole already at large distances, where their gravitational  wave emission 
is still negligible. Of these, only stars in almost radial orbits can
approach the horizon, where they can emit strong gravitational waves, as
well as being subject to tidal disruption. The rate of captures leading
to disruption may be even larger for He stars. Since a single periastron
passage near the tidal radius can disrupt the star, the detectable 
gravitational wave signal takes the form of a burst. However, the 
gravitational waves are expected also to be associated with X-ray/UV 
emission originating from the tidal compressional heating and disruption 
of the star. The detection of the photon flares provides the means for 
pinpointing the direction, occurrence time and possibly the redshift of 
the event, as well as for obtaining independent evidence for the 
existence and properties of the massive black hole. The temporal
coincidence of gravitational wave and electromagnetic signals could 
enhance the confidence level of the detection of both signals (e.g. Finn, 
Mohanty \& Romano 1999). Multi-window observations (gravitational and 
electromagnetic emissions) might allow us to set tighter constants on 
the properties of both the massive black hole and the star.

Here we study the gravitational and electromagnetic radiation from the 
tidal disruption of a main sequence star and a Helium star by a massive 
black hole. We use a three dimensional relativistic smoothed particle 
hydrodynamic (SPH) code (Laguna, Miller \& Zurek 1993), and investigate the 
gravitational wave and photon emission accompanying the tidal disruption 
event. In \S 2, we discuss the main features of the tidal disruption 
process and the photon flares expected during the disruption process. 
In \S 3 we provide a rough estimate of the characteristics of the 
gravitational waves. In \S 4 we discuss the numerical results obtained 
from relativistic SPH calculations around both Schwarzschild and 
maximally rotating Kerr black holes. In \S 5 the detectability of the
gravitational waves, and the constraints obtainable from photon and
gravitational wave observations are discussed. Conclusions and
discussions are given in \S 6. In the appendix, we discuss the 
shock formation during the tidal compression, using a relativistic Lagrangian 
code based on the Godunov method (Kobayashi, Piran \& Sari 1999).

\section{Stellar Disruption}

\begin{figure}
\plotone{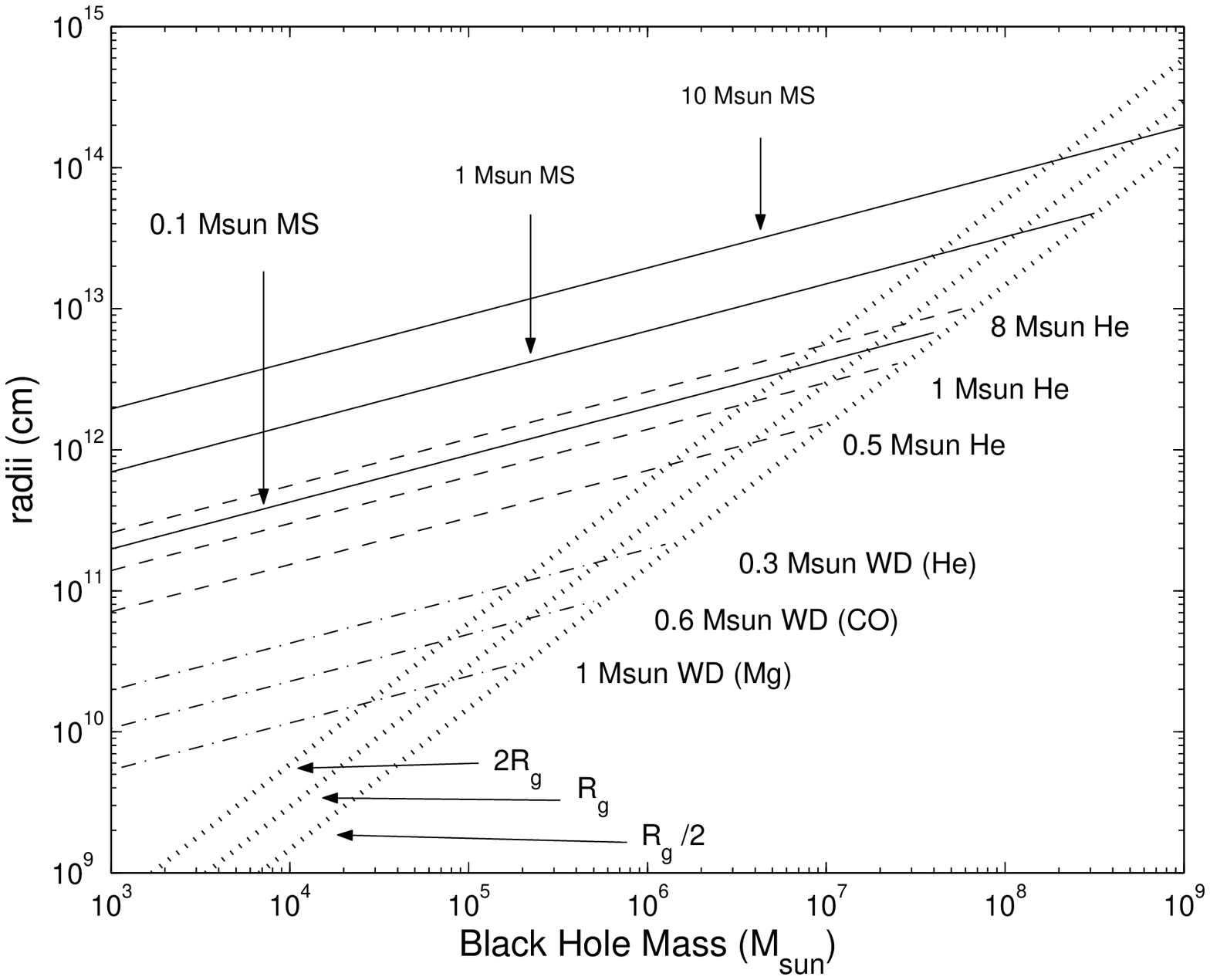}
\caption{
Tidal radius and event horizon scales for different stars as 
a function of the black hole mass, with notation $R_g=2GM/c^2$.
main sequence stars (MS), He stars (He) and white dwarfs (WD).
} 
\label{fig:rgrt}
  \end{figure}

The tidal disruption radius $R_t$ is the distance from a black hole at
which the tidal acceleration due to the black hole equals the self-gravity 
of the star. A simple estimate is
\begin{equation}
R_t \sim R_\ast \paren{\frac{M_{h}}{M_\ast}}^{1/3}
\sim 7 \times 10^{12} \paren{\frac{R_\ast}{R_\odot}}
\paren{\frac{M_\ast}{M_\odot}}^{-1/3}
\paren{\frac{M_{h}}{10^6 M_\odot}}^{1/3}
\mbox{cm}
\end{equation}
where $M_h$ is the mass of the black hole, $M_\ast$ and $R_\ast$ are
the mass and radius of the star, respectively. Thus, for a solar-type
star, the ratio between the tidal disruption radius $R_t$ and the event 
horizon scale $R_g=2GM_h/c^2 \sim 3 \times 10^{11} (M_h/10^6M_\odot)$cm is 
\begin{equation}
\frac{R_t}{R_g} \sim 24~
\paren{\frac{R_\ast}{R_{\odot}}}
\paren{\frac{M_\ast}{M_\odot}}^{-1/3}
\paren{\frac{M_h}{10^6M_\odot}}^{-2/3}.
\end{equation}
When the black hole mass is small enough ($M_h \siml 10^8 M_\odot$ for 
solar-type stars), the tidal disruption radius $R_t$ lies outside 
the event horizon, and the star is disrupted before falling in. 
In figure \ref{fig:rgrt}, we plot the tidal radii and the event horizon
scales for various types of stars as a function of the black hole mass.
Prompt disruption occurs when stars are scattered into highly eccentric
(loss cone) orbits with periastron distances $R_p < R_t$. The strength
of the tidal encounter is characterized by a penetration factor
$\beta_p$ defined as 
\begin{equation}
\beta_p=\frac{R_t}{R_p}.
\end{equation}

The tidal disruption of stars by massive black holes has been investigated 
by many authors both analytically and numerically (e.g. 
Carter \& Luminet 1982, 1983; Lacy, Townes \& Hollenbach 1982;
Nolthenius \& Katz 1982;  Luminet \& Marck 1985;
Rees 1988; Evans \& Kochanek 1989; Laguna et al. 1993; Khokhlov, Novikov
\& Pethick 1993; Marck, Lioure \& Bonazzola 1996; Loeb \& Ulmer 1997; Ulmer, 
Paczynski \& Goodman 1998; Ulmer 1999; Ayal, Livio \& Piran 2000; 
Ivanov \& Novikov 2001). 
During the dynamical evolution of a main sequence or a giant star,
several mechanisms related to the tidal force of a massive black hole
can lead to electromagnetic radiation from the star or the resulting debris.
Among these, the tidal compression and the fall-back of stellar debris
are likely to be most important in the context of the association with 
gravitational radiation.

(1) Tidal Compression: For a star penetrating well inside the tidal
radius, $\beta_p > 1$, the tidal disruption will be preceded by a 
short-lived phase of high compression of the star.  In the extreme case of
$\beta_p \gg 1$, the strong compression has been conjectured to cause a 
supernova-like thermonuclear detonation event (Carter \& Luminet 1982). 
However, even in the absence of a detonation, the thermal energy produced 
by the compression may lead to X-ray emission around the compression point,
as we discuss below.

A star penetrating deeply within the tidal radius of a massive black 
hole undergoes significant compression. Each section of the star is squeezed 
through a point of maximum compression at a fixed point on the 
star's orbit (see Fig 4 in Luminet \& Marck (1985)). This occurs on
a timescale comparable to the crossing time of the star through the 
compression point, 
\begin{equation}
\delta t \sim \frac{R_{\ast}}{v_p} \sim 10~ 
\paren{\frac{M_\ast}{M_\odot}}^{-1/6}
\paren{\frac{R_\ast}{R_\odot}}^{3/2}
\paren{\frac{M_h}{10^6M_\odot}}^{-1/3}
\mbox{sec}, 
\label{eq:tcross}
\end{equation}
where $v_p \sim c (R_g/R_t)^{1/2}$ is the orbital velocity at periastron.
The distortions $\Delta R_\ast/R_\ast \sim 1$ imply bulk flow velocities of 
order $R_\ast/\delta t \sim 10^{10}$ cm/s. Adiabatic compression alone 
can only increase the stellar surface and interior layer temperatures by a 
factor $\siml 10$, implying a sound speed  $c_s \sim (GM_\ast/R_\ast)^{1/2}
\sim 10^8$ cm/s. If this were the only heating, the adiabatic response time 
of the flow to the varying potential would be $\sim  R_\ast/c_s \sim 10^3$s, 
which is much longer than the dynamic crossing time (\ref{eq:tcross}). 
This implies a very large Mach number, and strongly suggests that shocks 
must occur.  Due to the complicated geometry and the large dynamic range, 
previous numerical experiments have not shown the development of shocks. 
Most calculations, however, are based on SPH methods, which are not well 
suited for studying shocks, but are resorted to in order to handle the 
complex geometry and potentials.  Our SPH calculations verify the magnitude 
of the large distortions mentioned above, on a timescale comparable to 
(\ref{eq:tcross}), and we are led to conclude that shocks are likely to occur. 
To verify the existence of shocks, we have performed additional numerical 
calculations on a one-dimensional model, using a different code specifically 
developed for handling relativistic shocks, a Lagrangian hydrodynamic code 
based on the Godunov method. The numerical results, discussed in the
Appendix, indicate that in deep penetration cases $\beta_p \gg 1$ a shock 
does indeed develop near the periastron and propagates through the star. 
Shocks are the only way to rapidly heat the gas, including the surface, to 
temperatures of order the virial value ($\sim$ keV), which are also 
approximately the temperatures indicated by our SPH calculations. 
Here we proceed on the assumption, substantiated by the above Lagrangian 
shock-capturing calculations, that shocks have occurred, and that the 
temperatures provided by the SPH calculations are approximately 
representative of the values resulting from the sudden shock heating 
occurring throughout the volume of the star.

The thermal energy $E_{th}$ produced by the shock heating can be roughly 
estimated as 
\begin{equation}
E_{th}=\frac{GM_\ast^2}{R_\ast}\sim 4\times10^{48} 
\paren{\frac{M_\ast}{M_\odot}}^2\paren{\frac{R_\ast}{R_\odot}}^{-1}
\mbox{erg}.
\end{equation}
The typical temperature of the shocked star is 
\begin{equation}
k_B T \sim \frac{G M_{\ast}m_p}{R_{\ast}}\sim  1 
\paren{\frac{M_\ast}{M_\odot}}\paren{\frac{R_\ast}{R_\odot}}^{-1}
\mbox{keV}
\label{eq:temp}
\end{equation}
where $m_p$ is the proton mass. The timescale $\delta t$ (equation 
[\ref{eq:tcross}]) gives a rough estimate for the heating time, while the 
cooling time from the surface layer is shorter. Hence the X-ray flare has 
a duration lasting approximately the crossing time (\ref{eq:tcross}).
Since the Thomson optical depth of the star is very large, 
$\tau_T \sim M_\ast \sigma_T/4\pi m_p R_\ast^2 \sim 
10^{10} (M_{\ast}/M_\odot)(R_{\ast}/R_{\odot})^{-2}$, most of the thermal 
energy remains trapped inside the star. However, a fraction of the thermal 
energy near the surface can be radiated promptly. The thermal energy
in the outermost Thomson mean free path is at least
$E_{rad,min} \sim (E_{th} /\tau_T ) \sim 4 \times 10^{38} 
(M_\ast /M_\odot ) (R_\ast / R_\odot ) \mbox{ergs}$, 
and an approximate lower bound on the X-ray flare luminosity 
is $L_{X,min} \simg 4 \times 10^{37} (M_\ast/M_\odot)^{7/6}
(R_\ast /R_\odot )^{-1/2}$ $(M_h / 10^6M_\odot )^{1/3} \mbox{ergs s}^{-1}$, 
of the order of the stellar Eddington luminosity. 
However, the total energy radiated during the flare (the X-ray fluence) 
can be larger than this, since shock heating of the surface continues over
the crossing time (\ref{eq:tcross}). The energy radiated from the surface
during this time (\ref{eq:tcross}) can be replenished by energy diffusing up
from deeper layers down to a depth $D\sim (c \delta t R_\ast /\tau_T )^{1/2}$ 
which is within a diffusion time comparable to the crossing time $\delta t$. 
An upper limit to the thermal energy or fluence which diffuses out from this 
depth is 
\begin{equation}
E_{rad,peak}\siml E_{th}\paren{\frac{c \delta t}{ R_\ast \tau_T}}^{1/2} \sim
10^{43} ~\paren{\frac{M_\ast }{M_\odot }}^{17/12} 
\paren{\frac{R_\ast}{R_\odot}}^{1/4}
\paren{\frac{M_h }{10^6 M_\odot}}^{-1/6}~ \hbox{erg}.
\label{eq:fluence}
\end{equation}
The corresponding upper limit on the flare luminosity is
\begin{equation}
L_{X,peak} \siml 10^{42}~ 
\paren{\frac{M_\ast }{M_\odot }}^{19/12} 
\paren{\frac{R_\ast}{R_\odot}}^{-5/4}
\paren{\frac{M_h }{10^6 M_\odot}}^{1/6} ~ {\rm erg}~{\rm s}^{-1}~.
\label{eq:lumax}
\end{equation}
After the peak, whose duration is approximately given by (\ref{eq:tcross}), 
the X-ray luminosity decays as $L \sim (c t /\tau R_\ast)^{1/2} (E_{th}/t) 
\propto t^{-1/2}$. Since the crossing time $\delta t$ is much
shorter than the pre-shock expansion time of the star $\sim R_\ast/c_s \sim
2\times 10^3 (M_\ast/M_\odot)^{-1/2}(R_\ast/R_\odot)^{3/2}$s, and at most
comparable to the post-shock expansion time, while the emission region is 
thin, $D \ll R_\ast$, we approximate the emission region as a roughly 
static gas slab to obtain eqs. (\ref{eq:fluence}),(\ref{eq:lumax}) and the 
initial decay scaling $L\propto t^{-1/2}$. A caveat concerning the prompt 
emission is that if a large amount of gas gets ripped off before the shock 
compression of the bulk of the star, the optical depth of patches of the 
ripped off gas could be large enough intercept the photons radiated along 
some directions, while allowing escape along others. This effect should be 
accounted for in the future.

(2) Matter fall-back: After the disruption, a fraction ($25\%\sim50\%$) of 
the material in the disrupted star remains gravitationally bound to the black
hole (Rees 1988; Ayal et al. 2000). It returns to the periastron, giving
a fall-back accretion luminosity evolving with time as $\sim t^{-5/3}$
(Rees 1988; Phinney 1989). The total energy radiated during the fall-back
phase can be less than that in the subsequent accretion disk phase. 
Nevertheless, since the fall-back phase is relatively short-lived 
($\sim$ years), it dominates the early luminosity of a disruption event
(after the prompt flare).
The time since disruption for this to happen and the peak luminosity are 
given by (e.g. Li, Narayan \& Menou 2002),
\begin{eqnarray}
\Delta t_{fallback} &\sim& 10 \paren{\frac{M_{h}}{10^6M_\odot}}
\paren{\frac{M_\ast}{M_\odot}}^{-1}\paren{\frac{R_\ast}{R_\odot}}^{3/2}
\mbox{days}, \\
L_{fallback,peak} &\sim& 10^{45} \paren{\frac{f}{0.1}}
\paren{\frac{M_{h}}{10^6M_\odot}}^{1/6}
\paren{\frac{M_\ast}{M_\odot}}^{7/3}
\paren{\frac{R_\ast}{R_\odot}}^{-5/2}
\mbox{erg s}^{-1}
\end{eqnarray}
where $f$ is the fraction of the stellar material falling back to
periastron.

3) Disk accretion: The fall-back material eventually settles into a disk,
which is accreted on a timescale of the order the viscous time $t_{visc}
\sim 10^3$ years (Li, Narayan \& Menou 2002). In the present paper, we
focus mainly on the prompt flash by the first of these three components.

\section{Gravitational Waves}

A main sequence star which penetrates the tidal radius will be
disrupted around the periastron, and the particles in the disrupted star
follow approximately independent Keplerian orbits. The resulting 
stretched-out debris trail is not compact enough to emit strong gravitational 
waves after leaving the first closest approach. The detectable gravitational 
wave signal will thus have a burst-like behavior, roughly characterized by 
an amplitude $h$, a frequency $f$ and a duration $\tau \sim 1/f$, where
\begin{eqnarray}
h &\sim& 
\frac{GM_\ast R_g}{c^2D R_p}
\sim  2 \times 10^{-22} \beta_p \paren{\frac{D}{10Mpc}}^{-1}
\paren{\frac{R_\ast}{R_\odot}}^{-1}
\paren{\frac{M_\ast}{M_\odot}}^{4/3}
\paren{\frac{M_h}{10^6M_\odot}}^{2/3}, \\
f &\sim& 
\paren{\frac{GM_h}{R_p^3}}^{1/2}
\sim 6\times10^{-4} \beta_p^{3/2}
\paren{\frac{M_\ast}{M_\odot}}^{1/2}
\paren{\frac{R_\ast}{R_\odot}}^{-3/2} \mbox{Hz}.
\end{eqnarray}
LISA will be able to detect gravitational waves of amplitude $h \simg
10^{-21}$ for burst sources in the frequency range  $f\sim 10^{-4}-10^{-1}$Hz
(Danzmann et al. 1998; Hughes 2002). Gravitational waves from
stellar disruption could thus be be detectable if $\beta_p \simg 1$ 
and the distance $D\siml 10$ Mpc. In the next section,
we compare these estimates of the gravitational radiation and photon 
emission against more detailed results obtained with a numerical SPH code.

The disruption process can affect the amplitude and wave form of the
gravitational waves. For comparison with the SPH results, it is useful to
estimate the radiation from a point particle with the same mass as the star. 
Since highly  eccentric elliptic orbits are observationally indistinguishable 
from an appropriate parabolic orbit, we will consider parabolic orbits for 
the SPH simulations and point particle calculations. The parabolic orbit for 
a point particle of mass $M_\ast$ around a Kerr black hole of mass $M_h$ and 
spin $a ~(|a| < R_g/2)$ is described by the following equations 
in the Kerr-Schild coordinates (e.g. Landau \& Lifshitz 1975;
Campanelli et al. 2001),
\begin{eqnarray}
\paren{\frac{d\varphi}{dt} -\frac{a}{\Delta}\frac{dR}{dt}}
\paren{1-\frac{R_gR}{c\Delta}\frac{dR}{dt}}^{-1}
&=& c \delta^{-1}
\bra{L-(L-a)\frac{R_g}{R}}, 
\label{eq:dpdt}\\
\paren{\frac{dR}{dt}}^2
\paren{1-\frac{R_gR}{c\Delta}\frac{dR}{dt}}^{-2}
&=& 
c^2 \paren{\frac{\Delta}{\delta}}^2
\frac{R_g}{R}
\paren{1-\frac{R_p}{R}}
\paren{1-\frac{R_0}{R}},
\label{eq:drdt}
\end{eqnarray}
where the orbit is assumed to be in the equatorial plane,
$\delta= R^2+a^2-a(L-a)(R_g/R)$, $\Delta=R^2-R_g R+a^2$, 
$c L$ is the angular momentum per unit mass, 
the radii $R_p$ (periastron: $R_p\ge R_0$) and $R_0$  are solutions of
$R^2-(L^2/R_g)R+(L-a)^2=0$. If the black hole is not rotating $(a=0)$, 
the smallest distance to which the test particle on a parabolic orbit
can approach and yet not be captured by the black hole is $R_p\sim 2R_g$.
If the black hole is maximally rotating, the closest approach possible 
in the equatorial plan is $R_p \sim R_g/2$ for a prograde case and 
$R_p \sim 2.9 R_g$ for a retrograde case.
For $a=0$, the angular momentum 
is given by $L^2=R_gR_p (1-R_g/R_p)^{-1}$. Since we can express the
angular momentum in terms of an impact parameter $b$ and an asymptotic
orbital velocity $v_\infty (\ll c)$ as $L=b (v_\infty/c)$,
the effective cross section subtended by the black hole for penetration
by the star to within a distance $R_p$ is given by 
$\sigma_p =\pi b^2 = \pi (v_\infty/c)^{-2} R_gR_p (1-R_g/R_p)^{-1}$. 
This has the noteworthy property of depending only linearly on $R_p 
~(>2R_g)$ (Carter \& Luminet 1983), multiplied by a factor 
$R_g (c/v_\infty)^2 \gg R_p$, in contrast with the purely geometric 
cross section which would be proportional to $R_p^2$. The fraction
of stars penetrating within the distance $R_p$ is directly proportional
to $\sigma_p$, so the fraction of stars suffering a prompt tidal disruption 
can be much larger than in an estimate based on the geometrical cross section.

\section{Numerical Simulations}

We employ a three dimensional relativistic SPH code (Laguna et al. 1993)
in which the hydrodynamics is calculated in a fixed Schwarzschild or
Kerr black hole background written in the ingoing Kerr-Schild form,
whose coordinates are horizon-penetrating and suitable for studying the
evolution of stars orbiting closely around the horizon or even falling
into it. The code has a variable smoothing length, and the time integration 
is performed with a second order Runge-Kutta integrator with adaptive time 
step (see Laguna et al. 1993 for details). For simplicity we use a polytropic
equation of state with adiabatic index 
$\Gamma=5/3$ to model the stars. The density profile is obtained by solving 
the Lane-Emden equations and the particles (typically $5000$ particles) 
are distributed over the stellar volume using the calculated profile. The 
initial position and velocity are chosen so that the star approaches the
massive black hole in a parabolic orbit with a periastron radius $R_p$. 
For all simulations the starting point is a radial distance of  $100 R_g$, 
and the mass of the black hole is assumed to be $10^6M_\odot$ 
(except for the calculations related to Fig \ref{fig:tmax}, in which
$10^5M_\odot$ black holes are also considered).

\subsection{Solar-Type Stars}

\begin{figure}
\plotone{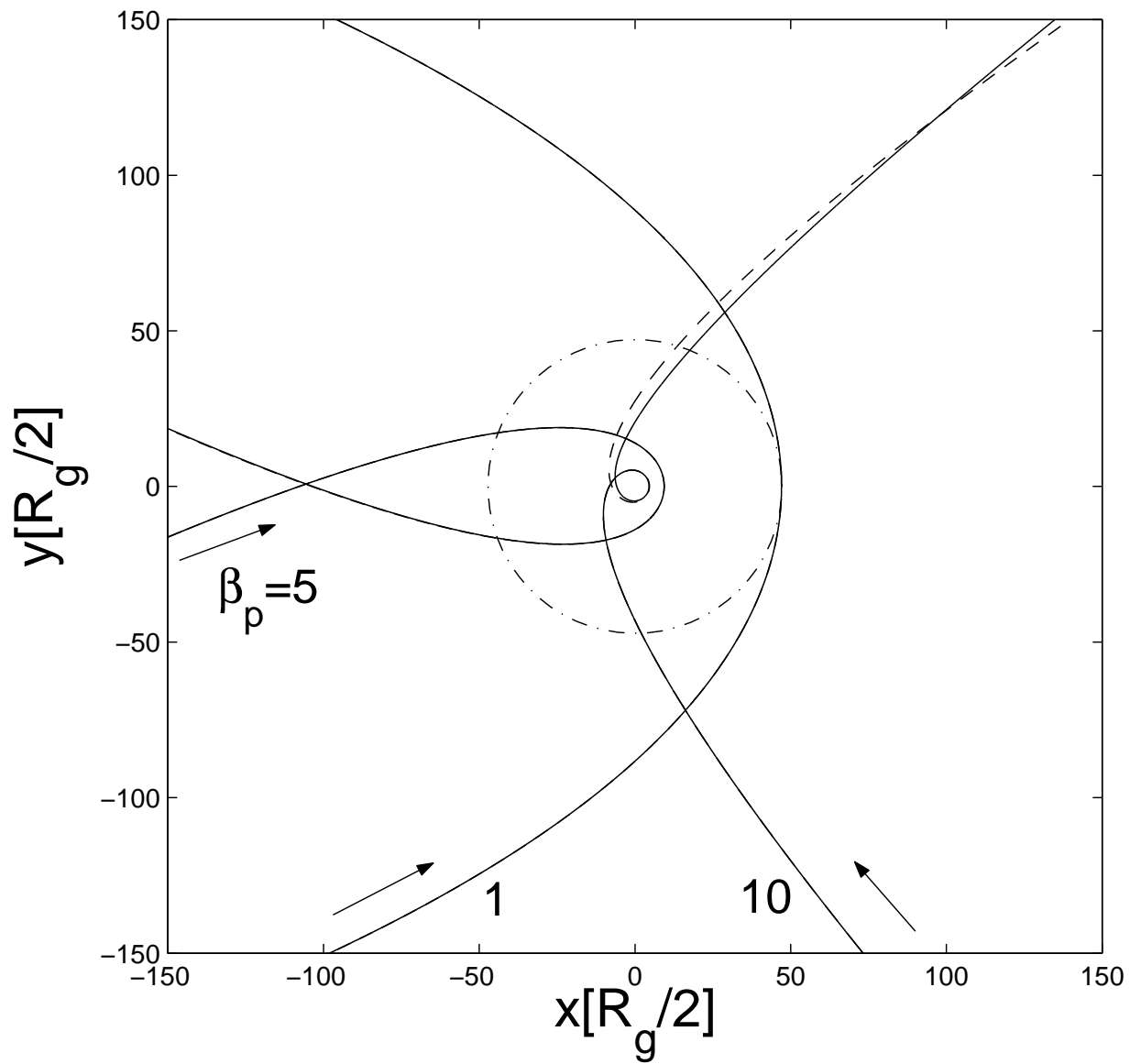}
\caption{Orbits of a solar type star with $\beta_p=1, 5$ or 10
around Schwarzschild black hole:
the orbit of the star's CM (solid) compared with the point particle
orbits (dashed). The dashed-dotted circle gives the tidal radius. 
The arrows indicate the motion of the stars.
$M_{h}=10^6M_{\odot}$ and $a=0$. The values of x and y coordinates are 
in unit of $R_g/2$.} 
\label{fig:orbits}
  \end{figure}

\begin{figure}
\plotone{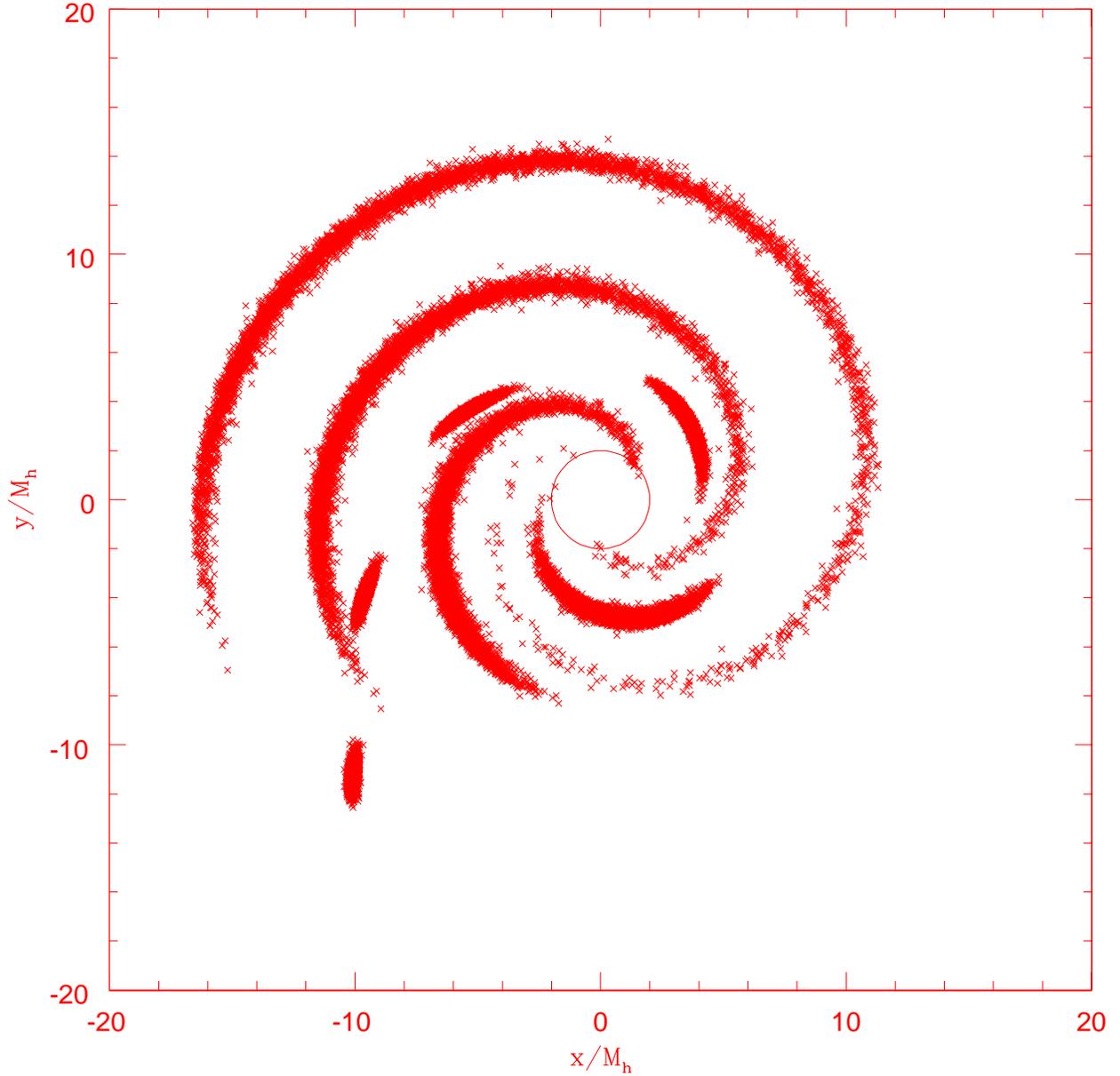}
\caption{Snapshots of disrupted star (SPH particle distribution):
   Eight frames at
   $t=-335, -236, -138, -39.7, 58.6, 157, 255$ and 353 sec
   are superposed. The time is relative to an origin at the instant of
   passage through the periastron of the orbit. $\beta_p=10$,
   $M_{h}=10^6M_{\odot}$ and $a=0$. The values of x and y coordinates are
   in units of $R_g/2=GM_h/c^2$. The center circle shows the Schwarzschild 
black hole horizon $R=R_g$.
  \label{fig:snapshot}}
  \end{figure}

We first consider the evolution of a solar-type star with
$R_\ast=R_\odot$ and $M_\ast=M_\odot$ in a parabolic orbit with 
penetration factors $\beta_p=(R_t/R_p)=1,5$ or $10$ around a Schwarzschild 
black hole with mass of $10^{6}M_\odot$. In Figure \ref{fig:orbits}, the 
solid lines indicate the orbits of the center of mass of the stars calculated 
with the SPH code, and the dashed lines indicate the orbits of the point
particle stars obtained with the geodesic equation. We note that 
the two estimates are indistinguishable in the cases of $\beta_p=1$ 
and 5. Because of the relativistic precession effect, the orbits with 
$\beta_p=5$ and 10 intersect with themselves. Especially for $\beta_p=10$, 
the periastron radius $R_p\sim 2.4 R_g$ is close to the critical value
$2R_g$ below which the star enters the black hole. The orbit tightly
winds around the black hole. A fraction of the kinetic energy is
dissipated during the break-up of the star, 
$\Delta E \sim GM_\ast^2/R_\ast$, but the effective deceleration is small, 
$\Delta  v/v_p \sim (GM_\ast R_t/c^2R_\ast R_g) \sim 10^{-4}$, where 
$v_p\sim c (R_g/R_t)^{1/2}$ is the orbital velocity at the periastron.
Therefore the center of mass of the distribution (solid lines) follows 
the orbit of the corresponding point particle (dashed lines). 
In the deep penetration case of $\beta_p=10$, the star is strongly disrupted 
after passing through the periastron. Snapshots of the particle distribution 
from the SPH simulations are shown in figure \ref{fig:snapshot}. When the
debris is located close to the critical radius $2R_g$, the orbital evolution 
of each debris element is sensitive to its own periastron. This causes the 
small difference between the SPH and point particle estimates.

\begin{figure}
\plotone{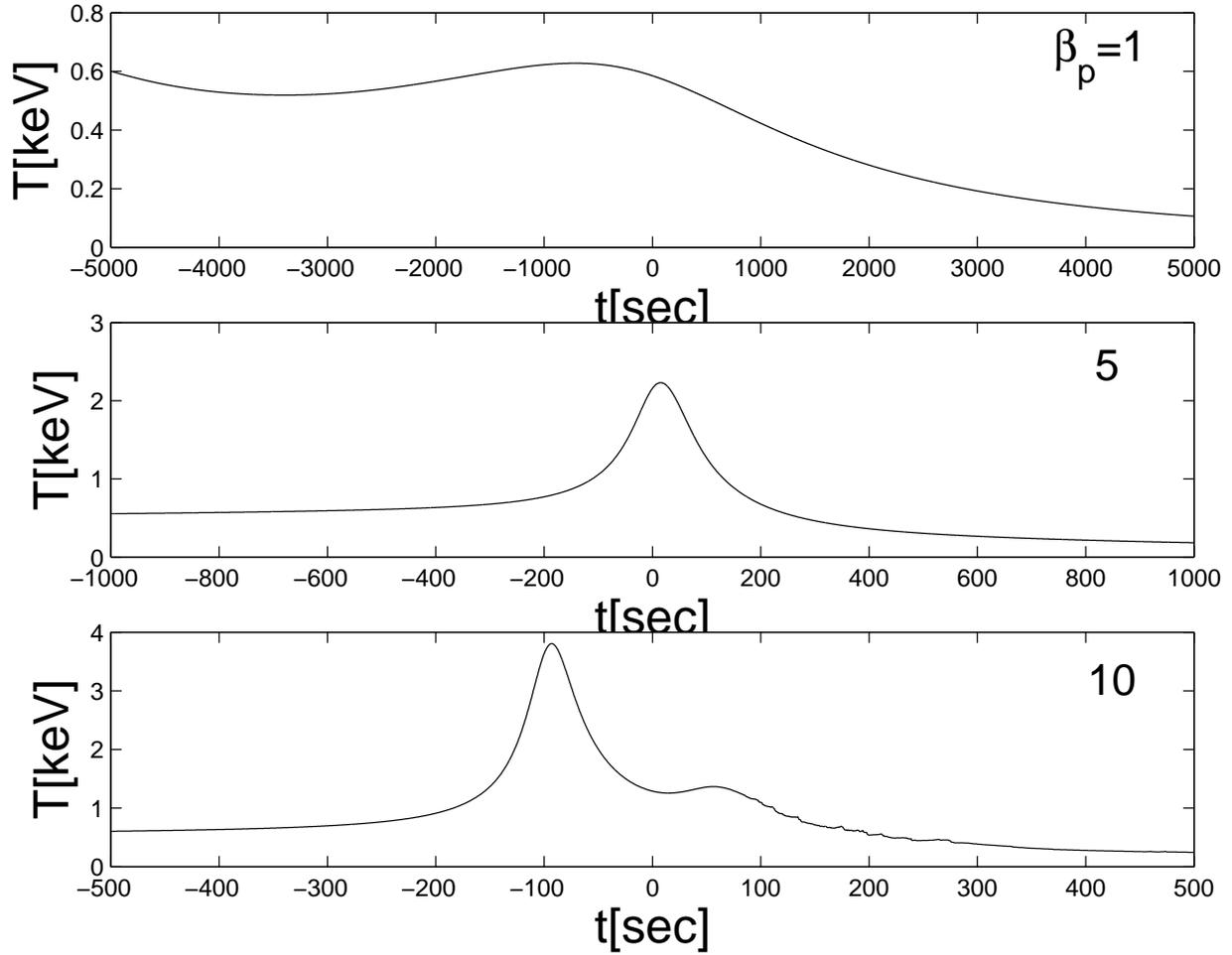}
\caption{Average temperature for solar-type star orbing Schwarzschild black 
hole. Time is relative to an origin at the instant of passage
through the periastron of the orbit.
  \label{fig:xsun}}
  \end{figure}

The evolution of the star's characteristic temperature $T$ is plotted in 
figure \ref{fig:xsun}. This temperature is the value averaged over all
particles.  It  provides a rough approximation to the thermal evolution 
of the star during the compressional heating and the subsequent re-expansion. 
The large optical depth of the star ensures that the photon diffusion 
time is much larger than any dynamical timescale, hence the adiabatic 
compressional heating and cooling of the SPH simulation provides a 
reasonable approximation to the thermal evolution.  The maximum temperatures 
for the values $\beta_p=1,5,10$ are $\paren{0.69, 2.2, 3.8}$ keV, and are 
consistent with the rough estimate in eq. (\ref{eq:temp}). 
The average 
temperature $T$ obtained from the SPH simulations, which in the unperturbed
star is weighted towards the central regions, becomes characteristic also
of the surface regions as shocks caused by the tidal compression reach the 
surface in a time comparable to the crossing time. This leads to photon
fluences and luminosities compatible with the estimates 
(\ref{eq:fluence},\ref{eq:lumax}).
The SPH average temperature depends rather weakly on the penetration 
parameter, roughly as $T \sim \beta_p$, and the corresponding emission 
is concentrated in the X-ray band. The first compression point 
is attained before the periastron. When the trajectory of the center of
mass intersects with itself within $R_t$, there are, at least, two
compression points, in agreement with Luminet and Marck (1985). 
Therefore, the temperature  has multiple peaks in the cases of 
$\beta_p \simg 7$, including $\beta_p=10$, as shown in figure \ref{fig:xsun} c. 
The second compression is less significant, but the the debris has a larger
surface area . The emission at the second peak might have 
a fluence similar to that of the first one.

\begin{figure}
\plotone{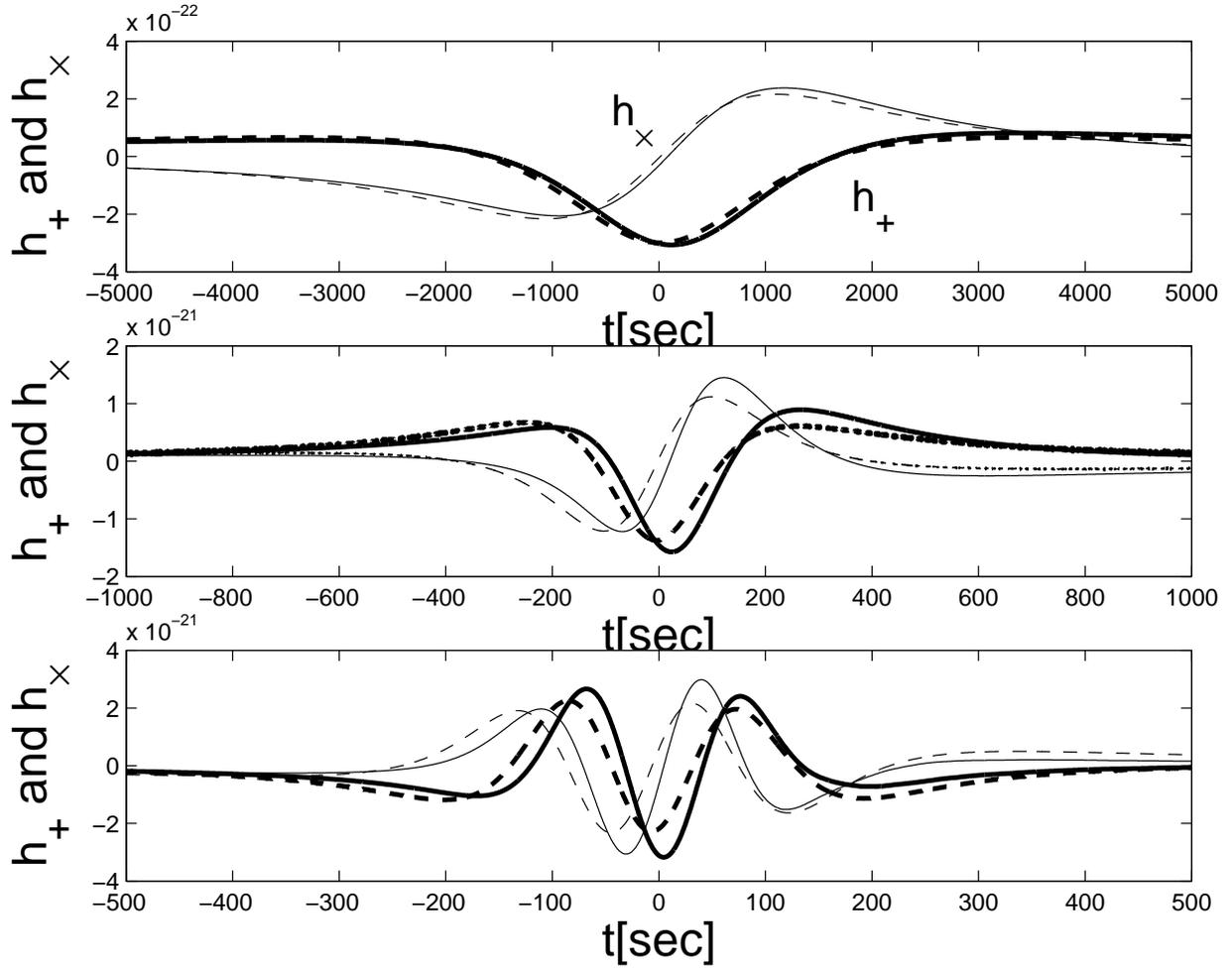}
\caption{Gravitational Waves from solar-type star orbiting around
  a Schwarzschild Black hole ($\beta_p=1,5$ or 10):
  SPH calculations (solid) and point particle estimates(dashed),
  $h_+$ (thick) and $h_\times$(thin)
  $M_h=10^6M_\odot$ and $D=20$Mpc.
  \label{fig:gwsun}}
  \end{figure}

We evaluate the gravitational wave emission using the quadrupole formula.
Although the velocity of the star at the periastron radius $v_p\sim
c(R_g/R_p)^{1/2}$ is close to speed of light, and the background
geometry is not flat, the quadrupole formula provides an approximate
estimate appropriate for our simple model. A more accurate treatment 
will be carried out in a future study. The wave forms of the gravitational 
waves emitted along the polar axis are shown in figure \ref{fig:gwsun}. 
The thick and thin lines give the two polarization components $h_+$ and 
$h_\times$, respectively. The solid lines indicate the SPH estimates 
(the sum of the emission from all particles), while the dashed lines
indicate the corresponding test particle estimates. The source is assumed 
to be at a distance $D=20$ Mpc. The trajectory in the case of $\beta_p=1$ goes 
halfway around the black hole, within a radius comparable to the periastron
radius (see figure \ref{fig:orbits}), resulting in approximately a single cycle of
gravitational wave emission. When the precession effect is larger
($\beta_p \gg 1$) a larger number of cycles are emitted, as one can see in
figure \ref{fig:gwsun} ($\beta_p=5$ and 10). Even in the case of
$\beta_p=10$, for which the star undergoes significant disruption (see
the distributions at $t>0$ in figure \ref{fig:snapshot}), the
wave form based on the point particle approximation agrees reasonably
well with the SPH results.

We have also evaluated the evolution of a ``dust'' star with the same 
initial conditions, i.e. one in which only gravity (the self gravity of 
the star and the background) is taken into account, and the hydrodynamic
module in the SPH code is turned off (zero-pressure). The resulting
gravitational wave emission is almost identical to that evaluated in the 
SPH calculation including hydrodynamics. This (and the agreement of the 
SPH and point particle estimates) suggests that the gravitational wave
signal during the disruption process is insensitive to the equation of
state and to the detailed inner structure of the star.  However, the 
detection of gravitational waves would provide information about the 
stellar orbit and the black hole.

\begin{figure}
\plotone{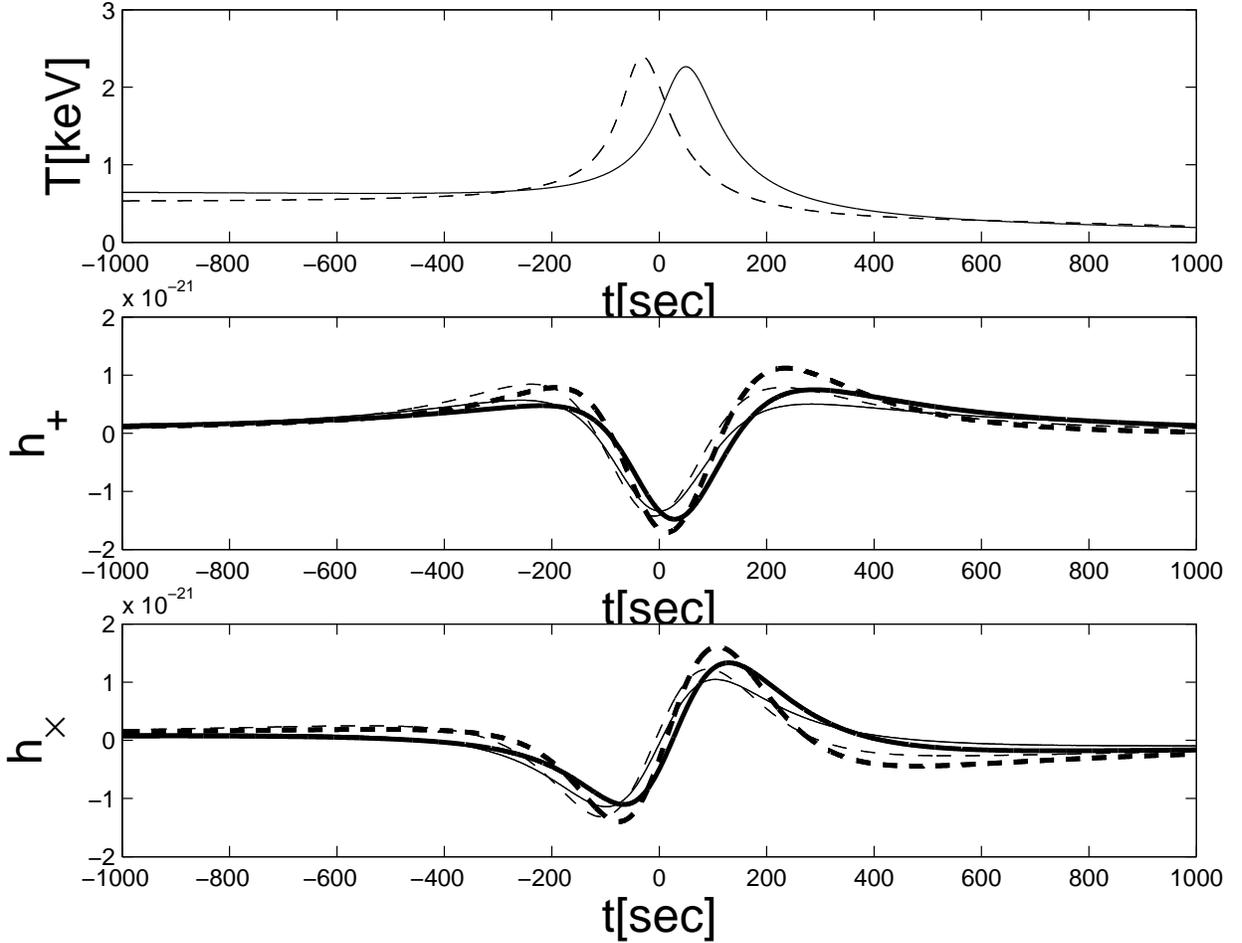}
\caption{Average temperature and gravitational radiation for solar-type stars
around a maximally rotating Kerr black hole in a $\beta_p=5$ orbit:
(a) Temperature (solid: prograde, dashed: retrograde) 
(b) $h_+$:  SPH estimates (thick) and point particle estimates (thin) 
(c) $h_\times$. Parameters are $M_h=10^6M_\odot$, $D=20$ Mpc.
  \label{fig:xgwspin}}
  \end{figure}

In order to see the effect of the black hole spin on the star's 
temperature and gravitational wave emission, we have evaluated
the evolution of a star orbiting around maximally rotating Kerr black
holes. The orbits are assumed to be in the equatorial plane, and a
direct (prograde) and retrograde orbits are considered. The 
penetration factor is fixed at $\beta_p=5$ for both cases. 
Figure \ref{fig:xgwspin} (a) shows the temperature of the star, while Figure 
\ref{fig:xgwspin} (b) and (c) show the two polarization components of the 
gravitational radiation along the polar axis. The temperature evolutions
are similar, but the gravitational wave signal depends on the spin of
the black hole. Thus, gravitational wave observations might allow a
determination of, or constraints on, the black hole spin.

\begin{figure}
\plotone{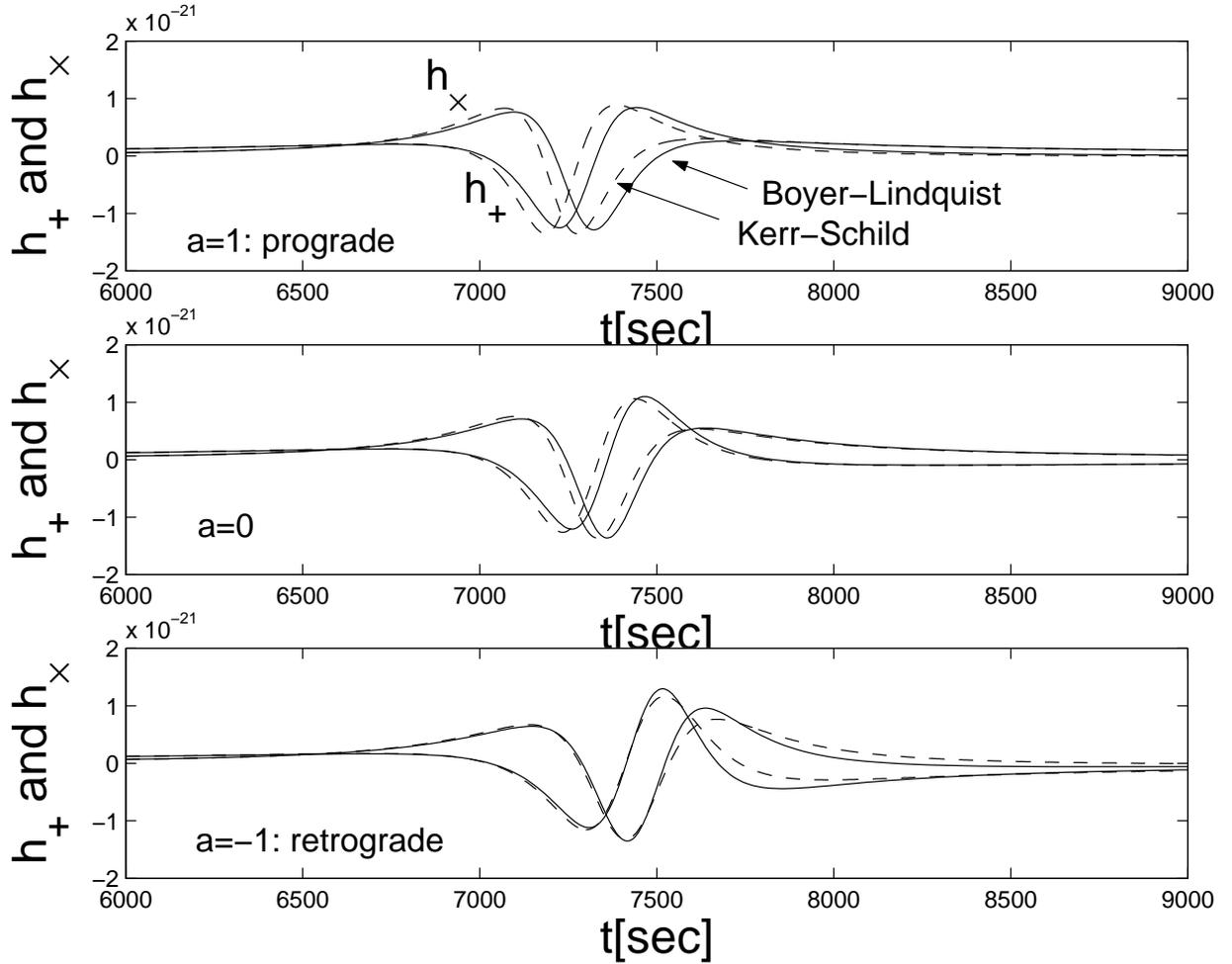}
\caption{Point particle estimates on gravitational 
waveforms $h_+$  and $h_\times$ in Boyer-Lindquist coordinates (solid) 
and Kerr-Schild coordinates (dashed).
Solar-type stars around a $10^6M_\sun$ black hole in a $\beta_p=5$ orbit.
(a) maximally rotating Kerr black hole (prograde)
(b) Schwarzschild black hole
(c) maximally rotating Kerr black hole (retrograde)
$D=20$ Mpc.
  \label{fig:KSBL}}
  \end{figure}

In our SPH code, the hydrodynamics is calculated in a Kerr black
hole background written in the (ingoing) Kerr-Schild form, which 
are horizon-penetrating and suitable for tracking the
evolution of stars orbiting closely around the 
horizon. However, the Kerr-Schild time is in a rotating frame. The
Boyer-Lindquist time which is in a fixed frame might be closer to what a
distant observer see. For comparison, we plot the wave signal results as
functions of Boyer-Lindquist and Kerr-Schild times in figure
\ref{fig:KSBL}, in which a solar-type star is assumed to be orbiting 
around a black hole with penetration factor $\beta_p=5$.
The two waveforms are almost identical for the Schwarzschild black hole 
case (fig \ref{fig:KSBL}b), but for Kerr cases, the difference could 
be noticeable (the difference of the strains at 
the same time is a few 10$\%$ level).

\subsection{He Stars}

The passage of giant stars within $\sim 10^2-10^4 R_g$ from a black hole
of mass $\sim 10^6M_\odot$, as well as tidal encounters between such stars 
in the nuclear region, can lead to the loss of the low density stellar 
envelope. The tidal stripping leaves the dense core (usually a He star) as
its end product. Within a few hundred $R_g$ from such a massive black
hole, the stellar population is likely to consist overwhelmingly of compact
stars including He stars. The rate of capture and tidal disruption of He 
stars by the massive black hole is expected to be higher than that of solar 
type stars.

\begin{figure}
\plotone{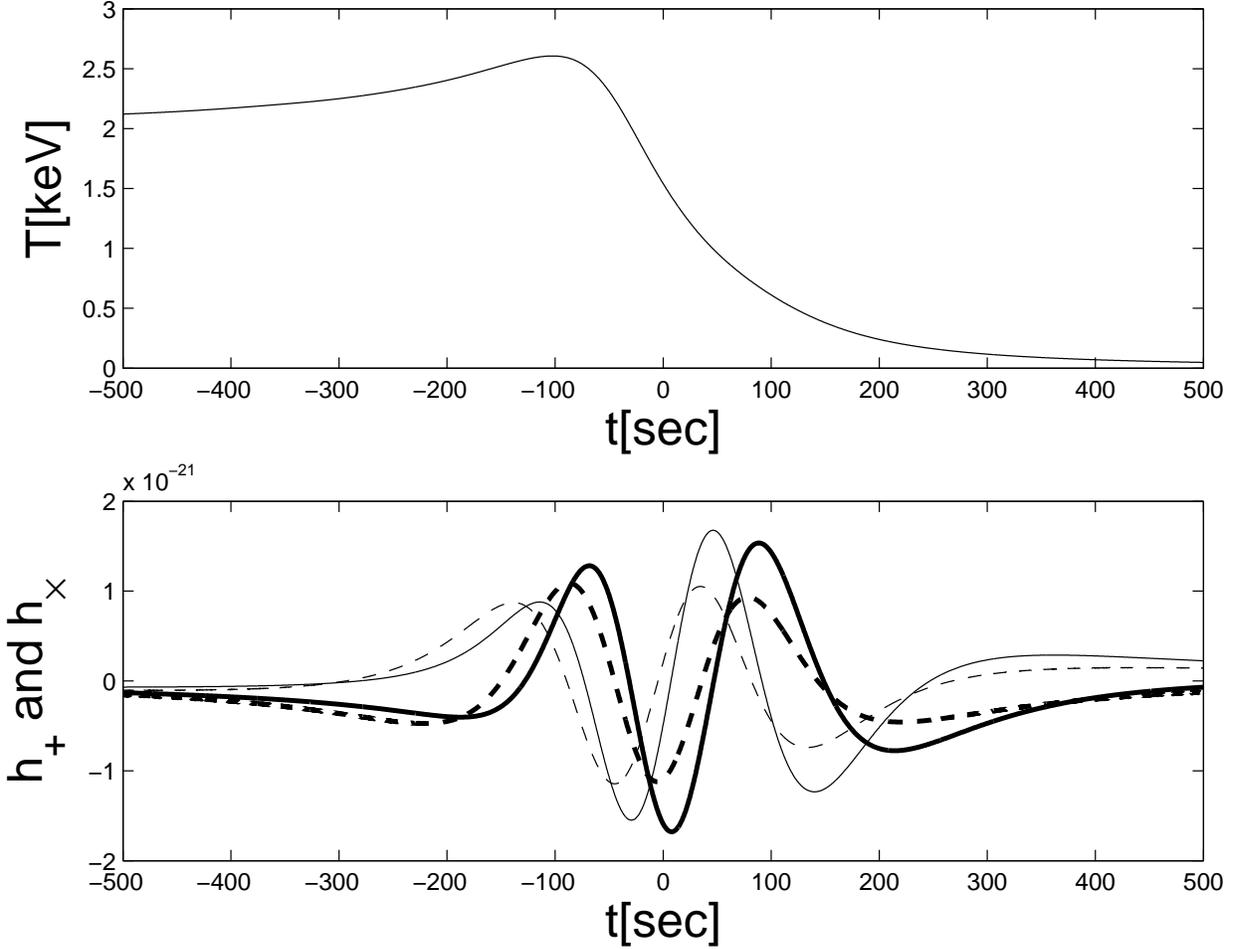}
\caption{Average temperature and gravitational radiation from a He star in a 
$\beta_p=1$ orbit around a Schwarzschild black hole of $10^6M_\odot$:
(a) temperature,
(b) gravitational wave strain: $h_+$ (thick) and $h_\times$ (thin), 
SPH estimates (solid) and point particle estimates (dashed). Parameters
are $D=20$ Mpc, $M_\ast=0.5 M_\odot$  and $R_\ast=0.08 R_\odot$.
  \label{fig:xgwhe}}
  \end{figure}

\begin{figure}
\plotone{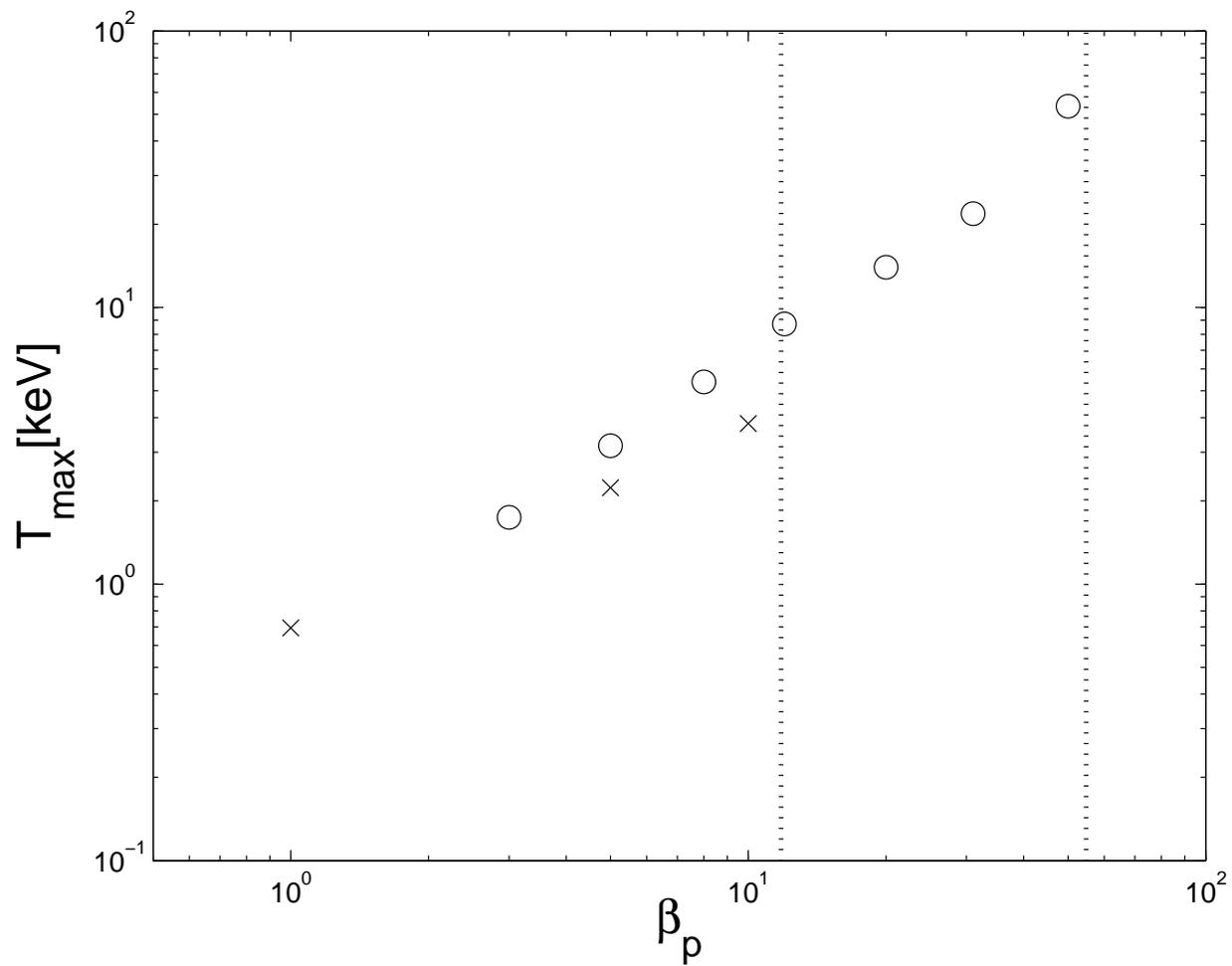}
\caption{Average temperature at the moment of maximum tidal compression vs.
   penetration factor
   $\beta_p$: a solar-type star orbiting around Schwarzschild black hole
   of $10^5M_\odot$ (circles) or  $10^6M_\odot$ (crosses). The vertical
   lines indicate the maximally allowed value of the penetration factor
   $\beta_{p,max}$. For a solar-type star, $\beta_{p,max} \sim 12$
   ($10^6 M_\odot$ BH) and $\sim 56$ ($10^5 M_\odot$ BH).
  \label{fig:tmax}}
  \end{figure}

We have studied the  evolution of He stars of mass
$M_\ast=0.5M_\odot$ and radius $R_\ast=0.08 R_\odot$  orbiting a
Schwarzschild black hole of  $10^6M_\odot$. The ratio of  $R_t$ to the
horizon scale $R_g$ is $R_t/R_g \sim 
2.4~(R_\ast/0.08 R_{\odot})(M_\ast/0.5 M_\odot)^{-1/3}(M_h/10^6M_\odot)^{-2/3}$. 
Since the critical periastron radius  $\sim 2R_g$, within which a star
is captured by the black hole, is close to the tidal radius $R_t$ 
the tidal heating of He stars in parabolic plunge orbits is relatively 
less severe than for solar-type stars, since only modest values of $\beta_p$ 
values are possible, for this hole $M_h=10^6 M_\odot$. However, a He star 
is more compact and in the unperturbed state has a larger central (and average) 
temperature than a solar star to begin with. In this case, numerical results 
for $\beta_p=1$ show that the tidal compression effect leads to larger average 
temperatures (figure [\ref{fig:xgwhe}]) than for the corresponding orbits in 
solar-type stars. The shock heated surface reaches temperatures of several keV. 
Thus, while the radius squared is down by $\sim 160$ the fourth power of the 
temperature is up by $\sim 300$ and the X-ray flare  has a peak luminosity 
comparable to equation (\ref{eq:lumax}), but with a harder spectrum.

Since the relativistic precession effect is large, the stellar orbit
winds tightly around the black hole, and multiple cycles of gravitational 
waves are emitted. The gravitational wave strain is, for the same $\beta_p$,
about one order of magnitude larger than for the corresponding solar-type 
stars, since the periastron is closer in. In this case also, the point 
particle approximation describes the SPH results reasonably well.


\section{Observational Prospects}

The tidal disruption rate depends on the mass of the black hole, the
stellar density, velocities in the galactic nucleus and the rate
at which radial loss cone orbits are depleted. The tidal disruption rate
of solar-type stars in samples of nearby galaxies was estimated as 
$10^{-6}-10^{-4}$ yr$^{-1}$ (Syer \& Ulmer 1999; Magorrian \& Tremaine 1999). 
The highest disruption rates (one star per $10^4$ yr) occur in faint 
$(L \siml 10^{10}L_\odot)$ galaxies.  The corresponding disruption 
rates for He stars are likely to be higher than these values.
Flares occur much less frequently in lager galaxies, partly because 
such galaxies have black holes with $M_h \simg 10^8M_\odot$ that 
swallow main-sequence stars whole, and partly because such galaxies 
are less centrally concentrated, meaning that their time-scales are 
much longer.

Observational constraints on disruption rates are uncertain.
A tidal disruption event, besides the prompt X-ray flash discussed here
is expected also to produce a long duration luminous photon flare in the 
fall-back phase (weeks to months after the stellar disruption), which 
may be looked for in all-sky soft X-ray surveys.  Komossa (2001) argued 
that the convincing detection of such a tidal disruption event would be the
observation of an event which fulfills the following three criteria:
(1) the event should be of finite duration (a 'flare'), (2) it should 
be very luminous, up to $\sim 10^{45}$erg s$^{-1}$ at maximum, and 
(3) it should reside in a galaxy that is otherwise perfectly non-active.
ROSAT all-Sky Survey (RASS) detected several candidates that fulfill the
criteria (summarized in Komossa 2001). 
Recent Chandra observations show that for three of the 
candidates the X-ray flux decline is consistent with the $t^{-5/3}$
decay predicted for the fall-back phase (Halpern, Gezari \& Komossa 2004). 
>From the statics of the RASS, 
Donley et al. (2002) obtained  a rate of $\sim 10^{-5}$ yr$^{-1}$,
which is consistent with the theoretical predictions.

As we have shown in \S 5, a gravitational wave burst with a large
amplitude is expected to be associated with a prompt X-ray flash 
when a solar-type or He star passes near the horizon of a 
massive black hole with mass of $10^6M_{\odot}$. This gravitational
signal might be detectable by LISA upto $D \sim 20$ Mpc. The Virgo
cluster consists of $\sim 2000$ galaxies, so one might detect such 
a gravitational wave burst once per 10-1000 yrs
\footnote{High values of the penetration factor $\beta_p$ are only 
possible for relatively low mass supermassive black holes 
$M_{h}\siml 10^7M_\odot$. Recent surveys of supermassive black hole 
candidates give somewhat larger mass estimates (a few $10^7M_\odot$; 
e.g. Kormendy \& Gebhardt 2001). If massive black holes in the Virgo 
cluster would predominantly have such larger masses, the event rates
of gravitational wave bursts associated with strong X-ray flashes could 
be lower than we calculated here. On the other hand, the dynamical
black hole mass estimates in our own Milky Way nucleus (Genzel et al. 2003;
Ghez et al. 2003) indicate that many, if not most, modest-sized galaxies
may have black holes in the mass range $\sim 10^6~M_\odot$, and these
may constitute the majority of galaxies}. 
The flux of the prompt X-ray flash from a source in Virgo is about
$\sim 10^{-10}$ erg cm$^{-2}$ s$^{-1}$, which for a typical duration 
of 10 s, which is detectable with the 23'x23' FOV Swift XRT detector, 
but falls one order of magnitude below the sensitivity of the Swift 
BAT detector, whose 2 steradian field of view and rapid slewing response 
would be key advantages. This flux is well above the sensitivity limits 
of the Chandra and XMM-Newton X-ray satellites, but their fields of view 
are narrow and slewing takes hours, so the chances are better for
detecting the longer (days to weeks) duration fall-back X-ray flare, 
which with the same spacecraft is detectable out to at least 200 Mpc.
The prompt X-ray flare, on the other hand, whose duration is of order 
$\sim 10$ s at a peak luminosity of $\sim 10^{42}$ erg/s, may require
a future wide field of view detector, a factor 30 more sensitive than
the BAT on Swift. The predicted time decay $\propto t^{-1/2}$ behavior
would help to distinguish these flares from type-I X-ray bursts. The 
detection would allow to observe the moment of the tidal disruption, 
and help confirm the detection of simultaneous gravitational waves.

The detection of the prompt X-ray emission could provide a constraint 
on the mass of the massive black hole involved in the tidal disruption
process. When a star passes near a massive black hole, the smallest 
distance to which the star in a parabolic or hyperbolic orbit can 
approach and yet not be captured by the black hole is $\sim 2R_g$ 
(Schwarzschild black hole). The ratio between this radius and the tidal 
radius $R_t$ gives the maximum value of the penetration factor $\beta_p$. 
($\beta_p \siml 12$ and $56$ for $10^6M_\odot$  and $10^5M_{\odot}$
black hole, respectively). The characteristic temperature of a solar-type 
star before the close passage is sub-keV. Our numerical results show that 
the tidal compression can increase this temperature by a factor of 
$\sim \beta_p$. In figure \ref{fig:tmax}, we plot the maximum 
temperature at the moment of the tidal compression as a function of
the penetration factor $\beta_p$. The crosses and circles indicate
the results for $10^{6}M_{\odot}$ and $10^{5}M_{\odot}$ black holes,
respectively. In the case of the $10^6 M_{\odot}$ black hole, the
temperature is about several keV at most. If an X-ray flare with much high
temperature is detected in a future mission, it might indicate that the
black hole involved in the tidal disruption process has a mass much smaller 
than $10^6M_\odot$ or that an intermediate mass black hole might be
harbored in the galaxy.

\section{Conclusions}

We have numerically investigated the evolution of solar-type and He-type
stars which are captured and disrupted by massive black holes with mass of
$10^{5-6}M_{\odot}$. We have shown that for both types of stars, plunge
orbits which approach the horizon produce gravitational wave bursts with 
large amplitudes, which may be detectable by LISA up to $D\sim 20$ Mpc. 
Even though the stars undergo significant disruption, the waveforms of the 
gravitational signal can be reasonably well approximated by a point particle 
approximation. We have also shown that the gravitational wave signal during 
the disruption process is largely insensitive to the equation of state and 
to the detailed inner structure of the star. For the same penetration
parameter $\beta_p$ (ratio of tidal to periastron radius), the waveforms 
of main-sequence and He stars differ considerably in shape, although
deep penetration (e.g. $\beta_p=10$) main-sequence waveforms have some
resemblance to critical penetration ($\beta_p=1$) He-star waveforms.
For large Kerr rotation parameters, the waveform shape and polarization 
is sensitive to the black-hole rotation and its sign.
In addition to long-term (days to years) X-ray flares, we predict also
a prompt X-ray flash, of luminosity $L_x\sim 10^{42}$ erg/s and duration 
$\sim 10$ s, associated with the gravitational wave bursts. The 
characteristic spectral temperature can give constraints on the black 
hole mass.

We thank M. Rees, S. Sigurdsson, T. Alexander, S. Mahadevan and the
referee for valuable comments. 
This work is supported by NASA NAG5-13286, NASA 
NAG5-10707, NSF AST 0098416, the Monell Foundation, the Merle Kingsley 
fund, and the Pennsylvania State University Center for Gravitational
Wave Physics, funded under cooperative agreement by NSF PHY 01-14375. 

\appendix
\section{Tidal Compression and Shocks: Numerical Study}

When a star penetrates within the tidal radius $R_t$ of a massive black
hole, the tidal acceleration begins to be dominant compared to
self-gravity. The star departs from hydrostatic equilibrium, and it
undergoes a significant compression. The compression eventually should be
halted and reversed by the build-up of pressure in a highly flattened 
pancake configuration (Carter \& Luminet 1982). For deep encounter cases 
(penetration factors $\beta_p \gg 1$), we expect that a shock occurs at
the end of the strong compression phase.

We investigate the shock formation process by means of a relativistic 
Lagrangian code based on the Godunov method with an exact Riemann solver 
(Kobayashi et al. 1999). Since the compression is governed by
the tidal force perpendicular to the orbital plane, we study a non 
self-gravitating gas slab in a varying gravitational potential. The 
effects of the tidal force can be simulated through
\begin{eqnarray}
\frac{d^2r}{dt^2} &=& -\frac{GM_h}{R^3}r, \\
R&=&\frac{2R_p}{1+\cos\theta}, \\
\frac{d\theta}{dt}&=&\paren{\frac{GM_h}{8R_p^3}}^{1/2}(1+\cos\theta)^2,
\end{eqnarray}
where $r$ is the distance from the orbital plane and $R(t)$ is the distance 
to the black hole for a Newtonian parabolic orbit. The hydrodynamic code
without the tidal force term has been tested in problems involving
ultrarelativistic flows, strong shocks and discontinuities (Kobayashi et 
al. 1999). In order to test the tidal effects in the code, we numerically
evolved a gas blob in a very rare and low pressure medium. For an initial
position $r_0$ and velocity $\dot r_0=0$ and a constant tidal field 
$R=$constant, the numerically obtained position of the blob is well 
described by the analytic solution $r(t)=r_0\cos kt$, where 
$k=(GM_h/R^3)^{1/2}$, except for the last moment $r\sim0$ when the 
build-up of pressure at the center decelerates the blob.

In the following calculations, we consider a stellar encounter characterized 
by $\beta_p=R_t/R_p=10$ and $R_t/R_g \sim 24$.  The latter corresponds to a
solar-type star and a $10^{6}M_\odot$ black hole. We assume 
a polytropic distribution with adiabatic index $\Gamma=5/3$ 
as the initial density and pressure distributions at $R(0)=R_t$. 
The central density is $\rho_c \simeq 5.9907 {\bar \rho} \simeq
(9M_\ast /2 \pi R_\ast^3)$. We express the density $\rho$ in units of
this central density $\rho_c$, and the pressure in units of $\rho_c c^2$.
The polytropic values become zero at the stellar surface $r=R_\ast$. For
numerical reasons, we evaluate the evolution of the fluid elements within
the radius $R_\ast^\prime (<R_\ast)$ within which $99\%$ of the star
mass is contained in the planar geometry model
($R_\ast^\prime\sim0.815R_\ast$). The pressure and density at 
$r=R_\ast^\prime$ are initially smaller than the central values by
factors of $\sim 90$ and $\sim 15$, respectively. We consider a uniform
ambient external medium. The pressure is assumed to be continuous at
$r=R_\ast^\prime$ and constant outside, while the outside density is 
$10^{-5}$ of the initial central density. All fluid elements in the star 
and ambient are initially at rest in the star's comoving frame. A 
reflection boundary condition is imposed at the center $r=0$.

\begin{figure}
\epsscale{0.85}
\plotone{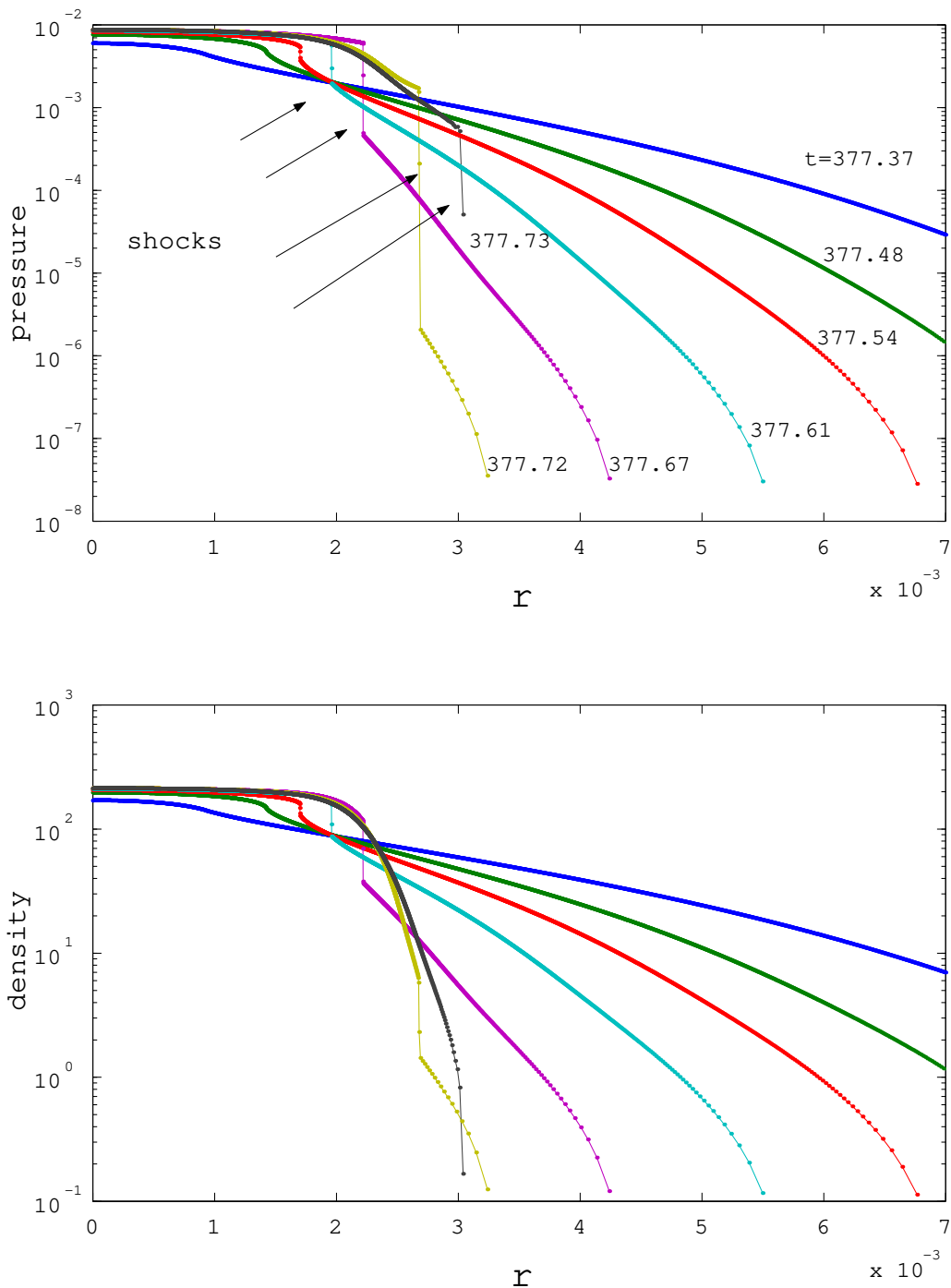}
\caption{Shock Formation: time evolution of the pressure and density 
in the tidal compression stage. The $r$-coordinate is the distance from 
the orbital plane in units of the initial stellar radius $R_\ast$. 
Time is in units of $R_\ast /c$, with $t=0$ set at the instant when
the star is at $R=R_t$. The density is normalized to an initial central 
density of unity, and the initial central polytropic pressure (for $c=1$) 
is $\sim 1.14 \times 10^{-6}$.
  \label{fig:shocks}}
  \end{figure}

Figure \ref{fig:shocks} shows the results for a 3000 zone mesh with equal
initial widths. In the figure we have plotted only the stellar region covered
by 2400 zones. After the star passes through the periastron at 
$t\sim 368.42 R_\ast/c$, the central pressure and density sharply
increase. At $t=377.37 R_\ast/c$, they are larger than the initial value
by a factor of $\sim 5200$ and $\sim 170$, respectively. At this point a 
shock starts to propagate from the center, and heats up the outer layers 
(see figure \ref{fig:shocks}). We performed an additional simulation in 
which the initial pressure and density are assumed to be uniform throughout 
the star, and equal to the central values of the polytropic distributions. 
This second calculation also shows a shock developing right after the 
periastron passage, which heats up the stellar material. This result 
indicates that the shock formation is rather insensitive to the choice 
of the stellar density distribution.


\end{document}